**Bound State Solutions of Exponential-Coshine Screened Coulomb plus Morse potential.**

Akpan N. Ikot[*1] and Oladunjoye A. Awoga[** 1] and Benedict I.Ita[***2]

[1] Theoretical Physics Group, Department of Physics, University of Uyo, Nigeria.

[2] Theoretical Quantum Chemistry Group, Department of Chemistry, University of Calabar, Nigeria.

[*] e-mail: ndemikot2005@yahoo.com

[**] e-mail: ola.awoga@yahoo.com

[***]e-mail: iserom2001@yahoo.com

Abstract

The analytical solutions of the Schrödinger equation with exponential coshine screened plus Morse (ECSM) potential are presented. The energy eigenvalues and the corresponding eigenfunctions are obtained for several values of screening parameters. We also present some numerical results for some selected diatomic molecules which are consistent with the values in the literature.

1. Introduction

The bound state solutions of the Schrödinger equation (SE) are only possible for some potentials of physical interest [1-4]. The exact or approximate solutions contain all the necessary information for the quantum system. In recent year, the problem of exact or approximate solution of the Schrödinger equation for a number of special potentials has been of great interest [5-7]. Different authors have obtained the exact solutions of the SE with some typical potentials using various methods. These potentials include the harmonic oscillator potential [8], Eckart potential [9], Woods-Saxon potential [10], Pseudoharmonic oscillator potential [11], ring-shaped kratzer-type potential [12], ring-shaped non-spherical oscillator potential [13], Rosen – Morse potential [14], Hulthen potential [15] etc. Different methods have also been employed in the study of those special potentials: Asymptotic Iteration Method (AIM) [16-17], Exact Method (EM) [18], Shifted $\frac{1}{N}$ Expansion [19], Super Symmetric Quantum Mechanics (SUSYM) [20], Tridiagonal J – Matrix [21], Algebraic Method [22] and Nikiforov – Uvarov (NU) Method[23].

Some of these potentials are known to play important roles in many fields of Physics such as molecular Physics, solid state Physics and chemical Physics [24]. One of such potential is the Morse potential and its vibrational energy levels are obtained using various techniques [25]. It has been reported that the potential energy function for the lowest electronic states of diatomic molecules can be expressed by the Morse Potential [25] as

$$V(r) = D_e\left[1 - e^{-\alpha(r-r_0)}\right]^2 \qquad (1)$$

Where $D_e$ is the dissociated energy, $\alpha$ is adjustable parameter, $r$ is the inter-atomic separation distance and $r_0$ is the distance from equilibrium position. Another well studied potential is the generalized exponential-cosine-screened coulomb (GECSC) potential defined as [26]

$$V(r) = -\frac{A}{r}e^{-\delta r}\cos(g\delta r) \qquad (2)$$

Where $A$ is the coupling strength constant, $\delta$ is the screening parameter, $g = 1$ is the cosine-screened coulomb constant. The special case of the GECSC potential known as the static screened coulomb (SSC) takes the form: $V(r) = -(\alpha Z e^2)exp(-\delta r)/r$ with $A = \alpha Z e^2$, $\alpha$ is the fine structure constant and Z is the atomic number and has been used in the description of the energy levels of light and heavy neutral atoms [26].

The exact solution for any $l \neq 0$ for the GECSC potential is still unknown. However, approximate methods [26] have been developed to evaluate the numerical and analytical values of its bound state.

For the purpose of this study, we investigate the Exponential Coshine Screened plus Morse (ECSPM) potential. This ECSPM potential is defined as

$$V(r) = D_e\left[e^{-2\delta r} - 2e^{-\delta r}\right] - D_e \cosh e^{-\delta r} \qquad (3)$$

Where $\delta$ is the screening parameter and $D_e$ is the dissociation energy. This potential can be used for the description of diatomic molecular vibrations [27]. They can also be useful in many branches of Physics for their bound and scattered problems [26]. Taking the first derivative of Eq. (3), that is, $\left(\frac{dV(r)}{dr}\right)_{r=r_0} = 0$, we get,

$$V(r_0) = \frac{(1 - D_e\delta/D_e)}{(1 + D_e\delta/2D_e)} - \frac{D_e\delta}{2} \qquad (4)$$

and

$$r_0 = \frac{1}{\delta} \ln\left(1 + \frac{D_e\delta}{2D_e}\right) \text{ for } \delta > 0, \qquad (5)$$

it can be observed that $r_0 = 1$ for $\delta << 1$ and $V(r_0) = -D_e - \frac{D_e\delta}{2}$. Similar observation can also be made for the following cases: (i) $D_e \to 0$ and (ii) $D_e\delta \to 0$ and the second derivative of the potential at $r = r_0$ determines force constant.

Satisfied with its performance through comparison with other methods, we decided to apply NU method to solve the ECSPM potential. Moreover, this work will show that the NU method can be a simple alternative way for computing the energy eigenvalues of diatomic molecules for this potential.

The organization of the paper is as follows: In section 2, the NU method for Schrödinger equation is outlined. In section 3, we derive the eigen values and eigen function of the ECSPM potential by NU method. In section 4, discussion and numerical results for some selected diatomic molecules are presented. We present the conclusion in section 5.

2. **Review of NU method**

The NU is based on solving the second order linear differential equation by reducing it to a generalized equation of hyper-geometric type. This method has been used to solve the Schrödinger, Klein –Gordon and Dirac equations for different kinds of potentials [28]. The second-order differential equation of the NU method has the form [23].

$$\psi''(r) + \frac{\bar{\tau}(s)}{\sigma(s)}\psi'(s) + \frac{\bar{\sigma}(s)}{\sigma^2(s)}\psi(s) = 0 \qquad (6)$$

Where $\sigma(s)$ and $\bar{\sigma}(s)$, are polynomials at most second degree and $\bar{\tau}(s)$ is a first-degree polynomial. In order to find a particular solution to Eq. (6), we use the common ansatz for the wave function as

$$\psi(s) = \varphi(s)\chi(s) \qquad (7)$$

It reduces Eq. (6) into an equation of hyper-geometric type

$$\sigma(s)\chi''(s) + \tau(s)\chi'(s) + \lambda\chi(s) = 0 \qquad (8)$$

and the other wave function $\varphi(s)$ is defined as a logarithmic derivative

$$\frac{\varphi'(s)}{\varphi(s)} = \frac{\pi(s)}{\sigma(s)} \qquad (9)$$

The solution of the hyper-geometric type function in Eq. (8) are given by the Rodriques relation

$$\chi_n(s) = \frac{B_n}{\rho(s)}\frac{d^n}{ds^n}(\sigma^n(s)\rho(s)) \qquad (10)$$

Where $B_n$ is a normalization constant and the weight function $\rho(s)$ must satisfy the condition

$$\frac{d}{ds}(\sigma\rho) = \tau\rho \qquad (11)$$

The function $\pi(s)$ and the parameter $\lambda$ required for the NU method are defined as follows:

$$\pi(s) = \frac{\sigma'-\bar{\tau}}{2} \pm \sqrt{\left(\frac{\sigma'-\bar{\tau}}{2}\right)^2 - \bar{\sigma} + k\sigma} \qquad (12)$$

$$\lambda = k + \pi' \qquad (13)$$

In order to find the value of $k$ in eq. (12), then the expression under the square root must be the square of the polynomials. Therefore, the new eigenvalue becomes

$$\lambda = \lambda_n = -n\tau' - \frac{n(n-1)}{2}\sigma'' \qquad (14)$$

where

$$\tau(s) = \bar{\tau} + 2\pi(s) \qquad (15)$$

whose derivative is negative. By comparing Eqs.(13) and (14), we obtain the energy eigen values.

## 3. Eigen values and Eigen function of the Schrödinger equation with ECSM Potential

The non-relativistic Schrödinger equation for s-wave is written as follows [22],

$$\left[\frac{-\hbar^2}{2\mu}\nabla^2 + V(r)\right]\psi(r,\theta,\varphi) = E\psi(r,\theta,\varphi) \qquad (16)$$

Where $V(r)$ is the ECSPM potential in Eq. (3). Using the transformation for the separation of variables for the wave function $\psi(r,\theta,\varphi) = {1}/{r} R(r)Y_{lm}(\theta,\varphi)$, we obtain the following sets of equations

$$\frac{d^2 R_{nl}}{dr^2} + \frac{2\mu}{\hbar^2}[E - V(r)]R_{nl}(r) = 0 \qquad (17)$$

$$\frac{d^2\Theta(\theta)}{d\theta^2} + \cot\theta \frac{d\Theta(\theta)}{d\theta} + \lambda\Theta(\theta) = 0 \qquad (18)$$

$$\frac{d^2\Phi(\varphi)}{d\varphi^2} + m_l^2\Phi(\varphi) = 0 \qquad (19)$$

Where $Y_{lm}(\theta,\varphi) = \Theta(\theta)\Phi(\varphi)$ represents the product of the solutions of Eqs. (18) and (19). The function $Y_{lm}(\theta,\varphi)$ is then known as the spherical harmonics and its solutions are well-known [29]. Equation (17) is the radial Schrödinger equation, and is the subject of investigation.

Substitution of Eq. (3) into Eq. (17), we obtain the radial wave equation as

$$\frac{d^2 R_{nl}}{dr^2} + \frac{\mu}{\hbar^2}\left[E_{nl} - D_e e^{-2\delta r} + 2D_e e^{-\delta r} + D_e \delta \cosh(\delta r)e^{-2\delta r}\right]R_{nl}(r) = 0 \qquad (20)$$

Defining a new variable $s = e^{-2\delta r}$ and substituting it into Eq. (20), we obtain the following equation

$$\frac{d^2 R_{nl}}{dr^2} + \frac{1}{s}\frac{dR_{nl}}{ds} + \frac{1}{s^2}[-\varepsilon^2 + \beta^2 s^2 + \gamma^2 s]R_{nl}(s) = 0 \qquad (21)$$

Where

$$\varepsilon^2 = -\frac{-2\mu}{\hbar^2 \delta^2}\left[E - \frac{D_e \delta}{2}\right] \qquad (22)$$

$$\beta^2 = \frac{-2\mu}{\hbar^2 \delta^2}\left(\frac{D_e \delta}{2} - D_e\right) \qquad (23)$$

$$\gamma^2 = \frac{4\mu D_e}{\hbar^2 \delta^2} \qquad (24)$$

In order to solve Eq. (21), we apply the NU method to it. By comparing Eq. (21) with Eq. (6), we obtain the following polynomials:

$$\bar{\tau} = 1, \quad \sigma(s) = s, \quad \bar{\sigma}(s) = \beta^2 s^2 + \gamma^2 s - \varepsilon^2 \qquad (25)$$

Substituting these polynomials into Eq. (12), we get the $\pi(s)$ function as

$$\pi(s) = \pm\sqrt{-\beta^2 s^2 + (k - \gamma^2)s + \varepsilon^2} \tag{26}$$

According to the basic requirement of the NU method, the expression in the square root of Eq. (26) must be the square of the polynomial. Thus, the solution of Eq. (26) gives the k – values as

$$k_\pm = \gamma^2 \pm 2i\beta\varepsilon \tag{27}$$

In this case, we find four possible functions for $\pi(s)$ as

$$\pi(s) = \pm \begin{cases} i\beta s + \varepsilon, for\ k = \gamma^2 + 2i\beta\varepsilon \\ i\beta s - \varepsilon, for\ k = \gamma^2 - 2i\beta\varepsilon \end{cases} \tag{28}$$

We now select

$$k = \gamma^2 - 2i\beta\varepsilon,\ \pi(s) = -i\beta s + \varepsilon \tag{29}$$

To obtain

$$\tau(s) = 1 - i2\beta s + 2\varepsilon,\ \tau'(s) = -2i\beta < 0 \tag{30}$$

Now, using Eqs. (13) and (14) and together with Eqs (29) and (30), we get

$$\lambda = \gamma^2 - 2i\beta\varepsilon - i\beta, \tag{30a}$$

$$\lambda_n = 2ni\beta,\ n = 0, 1, 2 \tag{30b}$$

Now taking $\lambda = \lambda_n$, we can solve Eqs (30a and 31b) to obtain the energy equation with the ECSPM potential for the Schrödinger equation as,

$$E_n = \frac{-\hbar^2\delta^2}{8\mu}\left[1 + 2n + \left(\frac{4D_e}{\frac{D_e\delta}{2} + D_e}\right)^2\right]^2 - \frac{D_e\delta}{2} \tag{31}$$

Where the screening parameter $\delta \neq 0$. Two special cases can be deduced from the energy spectrum in equation (31):

(i) Coshine Screened Potential (CSP): This is obtained when the dissociation energy $D_e \to 0$ and its energy eigen values becomes

$$E_n^{CSP} = \frac{-\hbar^2\delta^2}{12\mu}\left(n + \frac{1}{2}\right)^2 - \frac{D_e\delta}{2} \tag{32}$$

(ii) Morse Potential (MP) is obtained when the $D_e\delta \to 0$ so that its energy eigen values becomes

$$E_n^{MP} = \frac{-\hbar^2\delta^2}{2\mu}\left[n + \frac{1}{2} + 2D_e\right]^2 \tag{33}$$

Let us now calculate the wave function. The wave function of the system can be determined as follows: Substituting Eq. (30b) into Eq. (11), we obtain the explicit form of the weight function as

$$\rho(s) = s^{2\varepsilon} e^{-2i\beta s} \tag{34}$$

which gives the first part of the wave function of Eq. (10) as

$$\chi_n(s) = B_n L_n^{2\varepsilon}(2i\beta s) \tag{35}$$

where $B_n$ is the normalization constant and $L_n^p(x)$ is the associated Laguerre polynomials. Similarly, the second part of the wave function is obtained from Eq. (9) as

$$\varphi(s) = s^{\varepsilon} e^{-i\beta s} \tag{36}$$

Hence we find the unnormalized wave function expressed in terms of the Jacobi polynomials as

$$\psi_n(r) = N_n \frac{1}{r} \left(e^{-\delta r}\right)^{\varepsilon} e^{-i\beta r} L_n^{2\varepsilon}(2i\beta) e^{-\delta r} Y_{ln}(\theta, \varphi) \tag{37}$$

where $N_n$ is the normalization constant

## 4. Discussion and Numerical Results

The behaviour of the energy eigen spectrum of the SE with ECSPM potential is controlled by the screening parameter, $\delta \neq 0$. The two limiting cases of this quantum system yields:

(i) CSP, whose eigen values is given by Eq. (32) and its corresponding eigen function is

$$\psi_n^{CSP}(r) = N_n \frac{1}{\gamma} \left(e^{-\delta r}\right)^{\varepsilon} L_n^{2\varepsilon}\left(2i\beta' e^{-\delta r}\right) Y_{lm}(\theta, \varphi) \tag{37}$$

Where $\beta' = \sqrt{\frac{\mu z}{\hbar^2 \delta}}$

(ii) MP, the energy spectrum for this potential is as given in Eq. (33), and its corresponding wave function is,

$$\psi_n^{MP}(r) = N_n \frac{1}{\gamma} \left(e^{-\delta r}\right)^{\varepsilon'} L_n^{2\varepsilon'}\left(2i\beta' e^{-\delta r}\right) Y_{lm}(\theta, \varphi) \tag{39}$$

where

$$\varepsilon' = \frac{1}{\hbar\delta}\sqrt{-2\mu E}, \text{ and } \beta' = \frac{1}{\hbar\delta}\sqrt{-2\mu D_e}.$$

Now taking a value of physical parameter $D_e = 4.7446 Mev$, we show in Fig 1, the behaviour of the energy eigen values for different screened parameter $\delta = 0.000, 0.01 \text{ and } 0.015$ as a function of n, for the ECSM potential. Similar plot is shown in Fig 2 and Fig 3 for ECSP and MP. We also display in Fig 4 the variation of the ECSPM potential, ECSP and MP as a function of n with $\delta = 0.05$. The behaviour of this potential and its special cases is also plotted in Fig 5 and Fig 6 at $\delta = 0.01$ and $\delta = 0.05$ respectively. We also show in Fig 7 the variation of the potentials (ECSPM, ECSP and MP) as a function of $r$ for various values of the screening parameters ($\delta = 0.05, 0.01$ and $0.015$). As can be seen from Fig. 7 this potentials work well for small and large values of the screening parameters.

In order to verify the accuracy of Eq. (31), we calculate the energy eigen values twenty-three selected molecules: $H_2$, ZnH, CdH, CH, OH, HF, HCl, HBr, HgH, HI, $Li_2$, $Na_2$, $K_2$, $N_2$, $P_2$, $O_2$, SO, $Cl_2$, $Br_2$, $I_2$, ICl, CO and NO. The molecular constant of these molecules are given in Table 1 taken from ref.[30]. The energy states of these molecules are calculated and shown in Tables 1 and 2 for n = 0 to 10. These results hold for low and high values of $\delta$. However, for high values of $\delta$ the ECSPM prevails and the results become worst when $\delta$ are very small as can be seen in Tables 1 and 2 for $K_2$, $I_2$, $Br_2$ and ICl molecules whose screening parameters are very small (See Table 1) and compare with their energy states in Tables 2 and 3.

5. Conclusion

In this study, we investigate the bound state energies of the exponential coshine screened coulomb plus Morse potential using the NU method. We obtain the energy eigenvalues and the corresponding eigen functions for several values of screening parameters. The present results are physically reasonable and are listed in Tables 2 and 3. These results can be compared with those of the exponential cosine screened coulomb (ECSC) potential in Ref [26].

Acknowledgement

A. N. Ikot and O. A. Awoga wish to dedicate this work to their families for their love and care.

# CAPTION TO FIGURES

**Fig. 1:** The plot of energy eigenvalues of ECSPM as a function of **n** for different screening parameters $\delta=0.005 cm^{-1}$, $\delta=0.01 cm^{-1}$ and $\delta=0.15\ cm^{-1}$. The other parameter is $D_e$ = 4.746768 MeV .

**Fig. 2:** Variation of the eigenvalues of ECSP versus **n** for several screening parameters $\delta=0.005 cm^{-1}$, $\delta=0.01 cm^{-1}$ and $\delta=0.15\ cm^{-1}$ with $D_e$=4.746768 MeV .

**Fig. 3:** A plot of the bound states energy of MP versus **n** for the same paramaters as in Fig. 1 and Fig. 2.

**Fig. 4:** A plot of the three energies ECSPM, ECSP and EMP versus **n** for $\delta=0.005 cm^{-1}$ and $D_e$=4.746768 MeV.

**Fig. 5:** A plot of the three energies ECSPM, ECSP and EMP versus **n** for $\delta=0.01 cm^{-1}$ and $D_e$=4.746768 MeV.

**Fig. 6:** A plot of the three energies ECSPM, ECSP and EMP versus **n** for $\delta=0.015 cm^{-1}$ and $D_e$=4.746768 MeV.

**Fig. 7:** The variation of ECSPM, ECSP and MP with **r** for various values of the screening parameters $\delta=0.005 cm^{-1}$, $\delta=0.01 cm^{-1}$ and $\delta=0.015\ cm^{-1}$. Other parameters are $D_e$=4.746768 MeV.

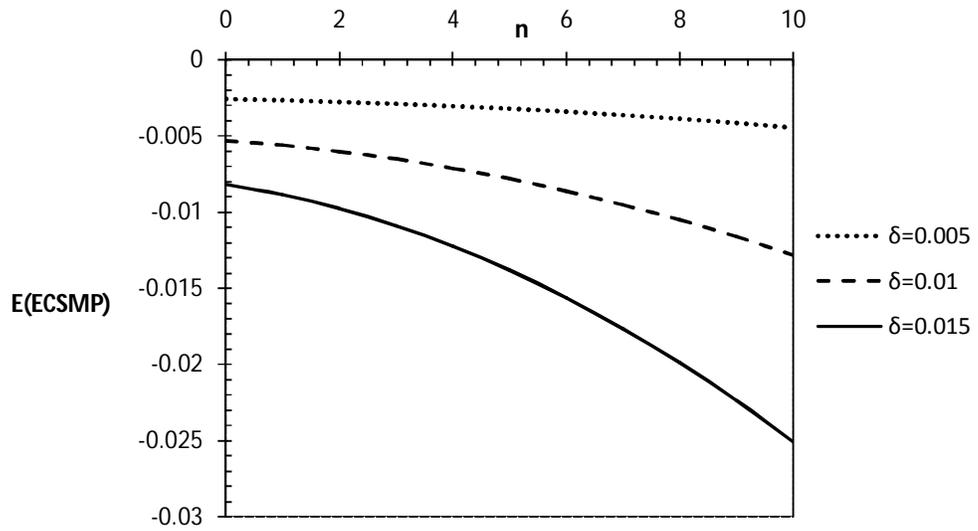

**Fig. 1**

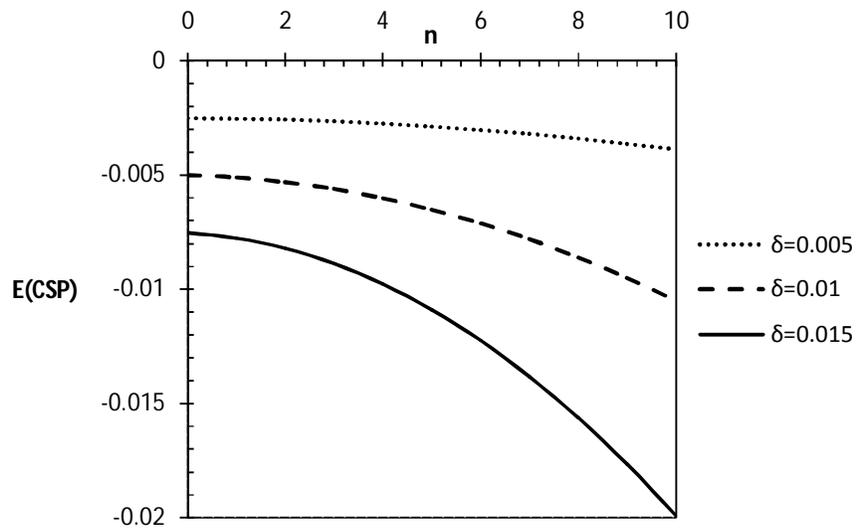

**Fig. 2**

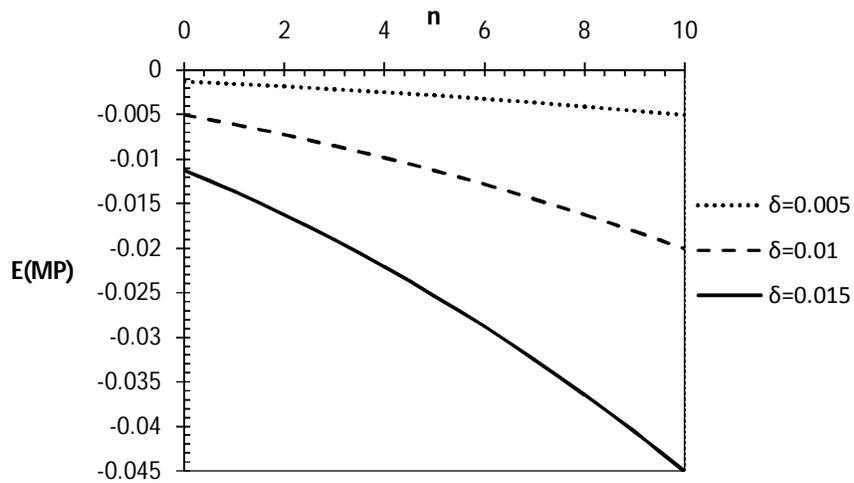

**Fig. 3**

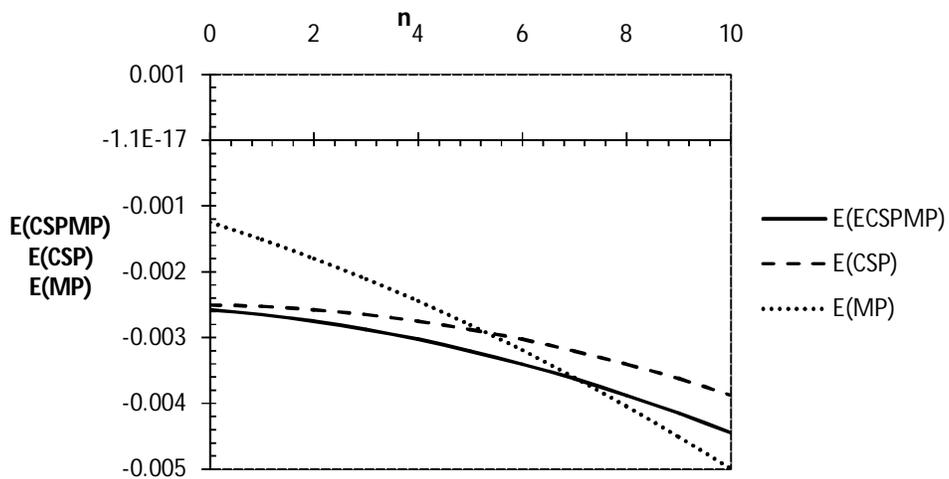

**Fig. 4**

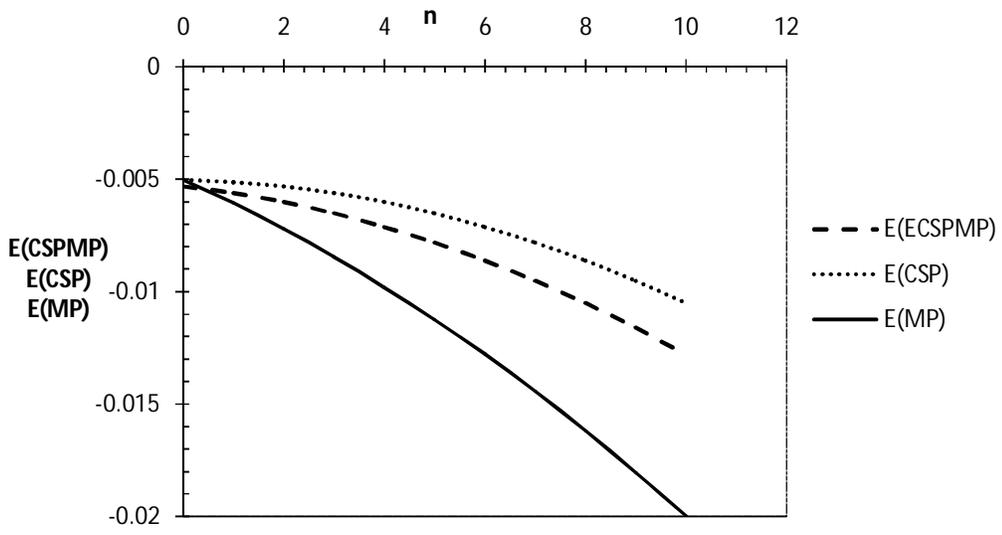

**Fig. 5**

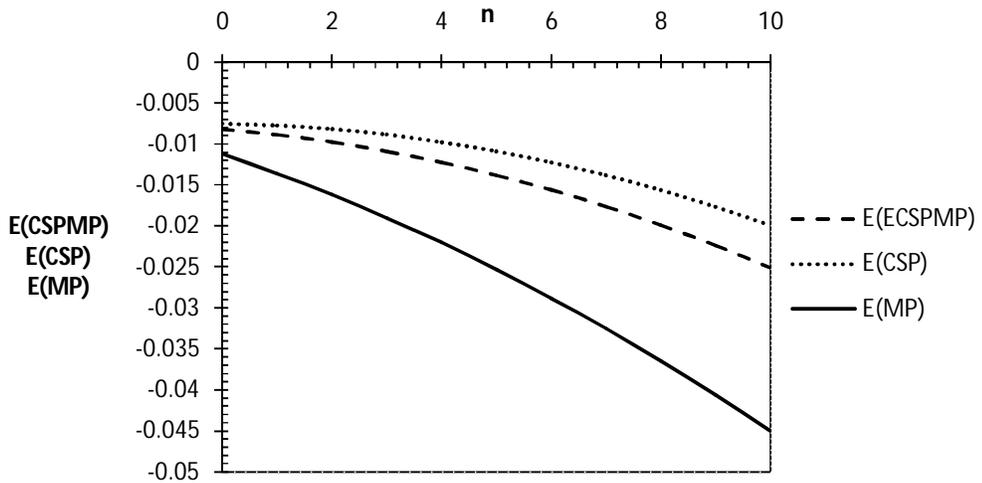

**Fig. 6**

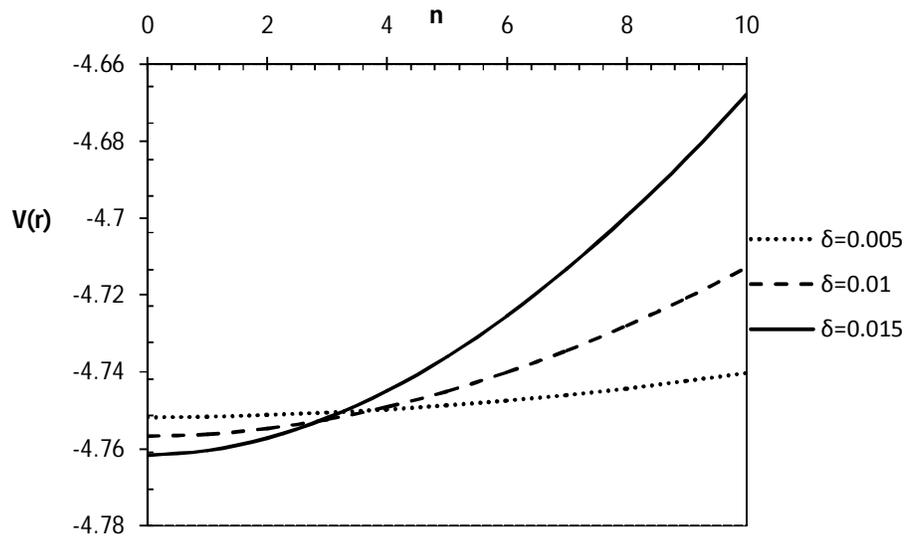

**Fig. 7**

Table 1: Molecular constants of some molecules [30]

| Molecules | μ(amu) | De(MeV) | $\delta(cm^{-1})$ |
|---|---|---|---|
| $H_2$ | 0.5041 | 4.746768 | 2.993 |
| ZnH | 0.9928 | 0.950352 | 0.25 |
| CdH | 0.9992 | 0.766272 | 0.218 |
| HgH | 1.0031 | 0.46136 | 0.312 |
| CH | 0.93 | 3.64032 | 0.534 |
| OH | 0.9484 | 4.58016 | 0.714 |
| HF | 0.9573 | 6.6456 | 0.77 |
| HCl | 0.9799 | 4.613856 | 0.3019 |
| HBr | 0.9956 | 3.916848 | 0.226 |
| HI | 1.0002 | 3.198 | 0.183 |
| $Li_2$ | 3.509 | 1.05144 | 0.00704 |
| $Na_2$ | 11.498 | 0.73944 | 0.00079 |
| $K_2$ | 19.488 | 0.51929 | 0.000219 |
| $N_2$ | 7.0038 | 7.516704 | 0.0187 |
| $P_2$ | 15.492 | 5.077488 | 0.00142 |
| $O_2$ | 8 | 5.211034 | 0.01579 |
| SO | 10.665 | 4.069728 | 0.00562 |
| $Cl_2$ | 17.489 | 2.509104 | 0.0017 |
| $Br_2$ | 39.958 | 1.99056 | 0.000275 |
| $I_2$ | 63.466 | 1.553136 | 0.000117 |
| ICl | 27.422 | 2.175264 | 0.000536 |
| CO | 6.8584 | 9.02616 | 0.01748 |
| NO | 7.4688 | 5.411952 | 0.0128 |

**Table 2: Energy eigenvalues of the molecules for n=0-5**

| Molecules/n | $E_n$ | | | | | |
|---|---|---|---|---|---|---|
| | 0 | 1 | 2 | 3 | 4 | 5 |
| $H_2$ | -13.091031 | -36.563890 | -77.807129 | -136.820749 | -213.604751 | -308.159133 |
| ZnH | -0.255007 | -0.417442 | -0.642830 | -0.931172 | -1.282466 | -1.696714 |
| CdH | -0.191028 | -0.315935 | -0.488403 | -0.708433 | -0.976026 | -1.291180 |
| HgH | -0.265405 | -0.507686 | -0.847010 | -1.283377 | -1.816787 | -2.447241 |
| CH | -1.439269 | -2.127900 | -3.123150 | -4.425019 | -6.033508 | -7.948616 |
| OH | -2.311257 | -3.432605 | -5.091485 | -7.287898 | -10.021844 | -13.293322 |
| HF | -3.295488 | -4.560583 | -6.445024 | -8.948812 | -12.071945 | -15.814425 |
| HCl | -0.884314 | -1.117758 | -1.444215 | -1.863685 | -2.376168 | -2.981664 |
| HBr | -0.557292 | -0.691421 | -0.876851 | -1.113584 | -1.401617 | -1.740953 |
| HI | -0.372085 | -0.461776 | -0.584948 | -0.741603 | -0.931740 | -1.155360 |
| $Li_2$ | -0.003745 | -0.003787 | -0.003843 | -0.003914 | -0.003998 | -0.004097 |
| $Na_2$ | -0.000292 | -0.000292 | -0.000293 | -0.000293 | -0.000293 | -0.000294 |
| $K_2$ | -0.000057 | -0.000057 | -0.000057 | -0.000057 | -0.000057 | -0.000057 |
| $N_2$ | -0.070433 | -0.070581 | -0.070778 | -0.071026 | -0.071324 | -0.071672 |
| $P_2$ | -0.003605 | -0.003606 | -0.003606 | -0.003607 | -0.003608 | -0.003609 |
| $O_2$ | -0.041236 | -0.041329 | -0.041452 | -0.041607 | -0.041793 | -0.042010 |
| SO | -0.011445 | -0.011454 | -0.011466 | -0.011481 | -0.011498 | -0.011519 |
| $Cl_2$ | -0.002133 | -0.002134 | -0.002134 | -0.002135 | -0.002136 | -0.002137 |
| $Br_2$ | -0.000274 | -0.000274 | -0.000274 | -0.000274 | -0.000274 | -0.000274 |
| $I_2$ | -0.000091 | -0.000091 | -0.000091 | -0.000091 | -0.000091 | -0.000091 |
| ICl | -0.000583 | -0.000583 | -0.000583 | -0.000583 | -0.000583 | -0.000583 |
| CO | -0.079024 | -0.079156 | -0.079333 | -0.079554 | -0.079820 | -0.080130 |
| NO | -0.034704 | -0.034769 | -0.034856 | -0.034965 | -0.035096 | -0.035249 |

**Table 3: Energy eigenvalues of the molecules for n=6-10**

| Molecules/n | $E_n$ | | | | |
|---|---|---|---|---|---|
| | 6 | 7 | 8 | 9 | 10 |
| $H_2$ | -420.483896 | -550.579039 | -698.444564 | -864.080469 | -1047.486756 |
| ZnH | -2.173916 | -2.714070 | -3.317178 | -3.983239 | -4.712253 |
| CdH | -1.653896 | -2.064175 | -2.522015 | -3.027418 | -3.580383 |
| HgH | -3.174737 | -3.999277 | -4.920860 | -5.939486 | -7.055156 |
| CH | -10.170344 | -12.698691 | -15.533657 | -18.675243 | -22.123448 |
| OH | -17.102333 | -21.448877 | -26.332953 | -31.754562 | -37.713704 |
| HF | -20.176250 | -25.157422 | -30.757940 | -36.977804 | -43.817013 |
| HCl | -3.680174 | -4.471697 | -5.356232 | -6.333782 | -7.404344 |
| HBr | -2.131590 | -2.573529 | -3.066770 | -3.611313 | -4.207157 |
| HI | -1.412462 | -1.703046 | -2.027112 | -2.384661 | -2.775692 |
| $Li_2$ | -0.004210 | -0.004337 | -0.004478 | -0.004633 | -0.004802 |
| $Na_2$ | -0.000294 | -0.000295 | -0.000295 | -0.000296 | -0.000296 |
| $K_2$ | -0.000057 | -0.000057 | -0.000057 | -0.000057 | -0.000057 |
| $N_2$ | -0.072069 | -0.072517 | -0.073014 | -0.073562 | -0.074159 |
| $P_2$ | -0.003610 | -0.003611 | -0.003612 | -0.003614 | -0.003615 |
| $O_2$ | -0.042259 | -0.042538 | -0.042849 | -0.043191 | -0.043564 |
| SO | -0.011543 | -0.011569 | -0.011599 | -0.011631 | -0.011667 |
| $Cl_2$ | -0.002139 | -0.002140 | -0.002142 | -0.002144 | -0.002146 |
| $Br_2$ | -0.000274 | -0.000274 | -0.000274 | -0.000274 | -0.000274 |
| $I_2$ | -0.000091 | -0.000091 | -0.000091 | -0.000091 | -0.000091 |
| ICl | -0.000583 | -0.000583 | -0.000584 | -0.000584 | -0.000584 |
| CO | -0.080485 | -0.080884 | -0.081328 | -0.081817 | -0.082350 |
| NO | -0.035424 | -0.035621 | -0.035840 | -0.036081 | -0.036343 |